\documentclass[manuscript, table, nonacm]{acmart}

\AtBeginDocument{%
  }

\usepackage{xcolor}
\usepackage{graphicx}
\usepackage{svg}
\usepackage{booktabs}
\usepackage{float}

\usepackage{amssymb, amsmath, amsfonts}
\usepackage{enumitem}
\usepackage{tabularx}
\usepackage{tabularray}
\usepackage{makecell}
\usepackage{url} 
\usepackage[color=pink!80]{todonotes}
\usepackage{color,soul}
\usepackage{tikz}
\usepackage{pifont}

\newcommand{\xmark}{\ding{55}}%

\usepackage{hyperref}

\makeatletter
\AtBeginDocument{%
  \def\doi#1{\url{https://doi.org/#1}}}
\makeatother

\begin{document}

\title[Identifying Periods of Cyclical Stress In-the-Wild]{Identifying Periods of Cyclical Stress in University Students Using Wearables In-the-Wild}

\author{Peter Neigel}
\email{peter.neigel@dfki.de}
\orcid{0000-0001-6336-2138}
\affiliation{%
  \institution{Osaka Metropolitan University}
  \city{Sakai}
  \state{Osaka}
  \country{Japan}
  \postcode{599-8531}
}
\affiliation{%
  \institution{German Research Center for Artificial Intelligence}
  \city{Kaiserslautern}
  \state{Rheinland-Pfalz}
  \country{Germany}
  \postcode{67663}
}

\author{Andrew Vargo}
\orcid{0000-0001-6605-0113}
\email{awv@omu.ac.jp}
\affiliation{%
  \institution{Osaka Metropolitan University}
  \city{Sakai}
  \state{Osaka}
  \country{Japan}
  \postcode{599-8531}
}

\author{Benjamin Tag}
\orcid{0000-0002-7831-2632}
\email{Benjamin.Tag@monash.edu}
\affiliation{%
  \institution{Monash University}
  \city{Melbourne}
  \state{Victoria}
  \country{Australia}
  \postcode{3800}
}

\author{Koichi Kise}
\orcid{0000-0001-5779-6968}
\email{kise@omu.ac.jp}
\affiliation{%
  \institution{Osaka Metropolitan University}
  \city{Sakai}
  \state{Osaka}
  \country{Japan}
  \postcode{599-8531}
}

\renewcommand{\shortauthors}{Neigel et al.}

\begin{abstract}

\sethlcolor{cyan}
University students encounter various forms of stress during their academic journey, including cyclical stress associated with final exams. Supporting their well-being means helping them manage their stress levels. In this study, we used a wearable health-tracking ring on a cohort of 103 Japanese university students for up to 28 months in the wild. The study aimed to investigate whether group-wide biomarkers of stress can be identified in a sample having similar daily schedules and whether these occurrences can be pinpointed to specific periods of the academic year. We found population-wide increased stress markers during exams, New Year's, and job hunting season, a Japanese job market peculiarity. Our results highlight the available potential of unobtrusive, in-situ detection of the current mental state of university student populations using off-the-shelf wearables from noisy data, with significant implications for the well-being of the users. Our approach and method of analysis allows for monitoring the student body's stress level without singling out individuals and therefore represents a privacy-preserving method. This way, new and sudden stress increases can be recognized, which can help identify the stressor and inform the design and introduction of counter measures.

\end{abstract}


\maketitle

\section*{Abbreviations}
HR - \textbf{H}eart Rate, HRV - \textbf{H}eart \textbf{R}ate \textbf{V}ariability, ANS - \textbf{A}utonomic \textbf{N}ervous \textbf{S}ystem, PNS - \textbf{P}arasympathetic \textbf{N}ervous \textbf{S}ystem, SNS -  \textbf{S}ympathetic \textbf{N}ervous \textbf{S}ystem, LF - \textbf{L}ow \textbf{F}requency Power Band of HRV, HF - \textbf{H}igh \textbf{F}requency Power Band of HRV, BPM - \textbf{B}eats \textbf{P}er \textbf{M}inute, ECG - \textbf{E}lectro \textbf{C}ardio\textbf{g}ram, SpO2 - \textbf{S}aturation of \textbf{P}eripheral \textbf{O}xygen

\section{Introduction}\label{sec:intro}

In humans, "stress" describes various physiological phenomena and responses to a stressor. These stressors can be acute, such as a presentation in front of a large audience or a life-threatening situation, or chronic, such as a stressful everyday job or excessive worrying. Stress, and stressful situations, can further occur in rather regular cycles. For example, if someone found the New Year's festivities the most stressful time of the year, there would be a one-year cycle for repeating stress. In this work, we look at university students who used wearable smart rings in the wild to analyze whether we could identify cyclical stress based on the timing of different periods of the academic year. This analysis takes place without adjusting for any other individual source of personal, work-related or other non-academic stress, meaning that classification is possible even with significant noise. 

The fight-or-flight hypothesis postulates that the physiological changes in the body in response to a stressor have the purpose of mobilizing all available energy resources, thereby increasing the individual's chances of either fighting or escaping the threat. A useful response in the face of short-term stressors, modern-day work-life and -load elicits these responses in many people routinely, however. When stress becomes regular and chronic, it can have serious effects on people's mental well-being, as well as causing actual physiological harm to the body. It is estimated that 80\% of people experience stress at work \cite{salleh_LifeEventStress_2008} and roughly half of the work-related illnesses are related to stress~\cite{europeancomission_CommissionAsksWorkers_2004}. The effects of permanent stress include worsened sleep~\cite{kim_EffectPsychosocialStress_2007}, anxiety and depression~\cite{richter-levin_HowCouldStress_2018}, headaches~\cite{stubberud_ThereCausalRelationship_2021}, weight gain~\cite{scott_EffectsChronicSocial_2012}, digestive problems~\cite{madison_StressDepressionDiet_2019}, problems with memory and focus~\cite{klier_StressLongtermMemory_}, muscle tension~\cite{zielinski_RelationshipStressMasticatory_2021}, and pain, and can, in severe cases, lead to heart disease, heart attacks~\cite{rozanski_ImpactPsychologicalFactors_1999}, high blood pressure~\cite{laitinen_SympathovagalBalanceMajor_1999}, and strokes~\cite{reddin_AssociationPsychosocialStress_2022}. When the body perceives a situation as threatening, the balance of activity in the ANS shifts away from the Parasympathetic Nervous System (PNS) towards the Sympathetic Nervous System (SNS), dilating pupils, increasing Heart Rate (HR) and forcing vasoconstriction. Heart Rate Variability (HRV), which is the fluctuation of the length between heartbeats, is a strong indicator of ANS balance (\cite{shaffer_OverviewHeartRate_2017} and section \ref{relwork}). It is usually calculated from the tachogram derived from Electrocardiogram (ECG) signals by analyzing QRS-complexes and obtaining the variance in the distance between two R-R peaks. Especially the LF/HF power ratio in the frequency domain analysis of HRV is indicative of ANS activity. Since these measurements require precise medical-grade equipment, a comprehensive detection of these physiological markers was not obtainable "in-the-wild" for most of the past. However, the recent ubiquity of wearable health devices such as smart rings or watches, with their capability of unobtrusively measuring physiological data like HR, HRV, body temperature, breathing rate, etc., over prolonged periods of time, makes this attainable.

In this work, we analyze the physiological data obtained from a wearable health tracker on a cohort of students in a Japanese university for group-wide changes in stress biomarkers. Usage of the device was not enforced or guided, resulting in natural in-the-wild data that is messy and has gaps. We are able to find heightened waking HR and maximum waking HR during exams, New Year's, and spring break when job hunting usually takes place. Sleep HR, sleep HRV, activity patterns, and sleep phases further substantiate the indication of stress-related effects.

The results of this study are promising for a number of reasons. First, it indicates that identifying stressful periods in a cohort in-the-wild is possible, which means that the results could be applied in education and the workplace for managing cyclical stress. This could be especially important for identifying and designing interventions where participants cannot self-report stress. In addition, the results represent an important opportunity for the HCI community to build information interfaces that allow users to better understand and prepare for cyclical stress.

The contribution of this work is to lay out the possibility of detecting group-wide changes in biomarkers indicating stress with a sample of a cohort wearing wearable health trackers without infringing on individual privacy or enforcing rules or guidelines on the usage of the device.

\section{Related Work}\label{relwork}

There is ample literature about the detection of stress and its effects on physiological measurements in laboratory settings with controlled stressors as well as for individuals in non-restricted environments using multiple sensors, including wearable devices, or even contextual data like phone activity and weather.

In contrast, the literature corpus dealing with (sub-)population-wide indicators of stress in rigid schedules is much smaller.

\subsection{University Environments}

\cite{can_StressDetectionDaily_2019, hickey_SmartDevicesWearable_2021} survey papers dealing with stress detection in daily life using smart-phones and wearable sensors. While the bulk of the analyzed works deals with laboratory tests and restricted environments, some papers dealing with university environments and completely unrestricted daily life are surveyed. In the papers dealing with university environments, most analyze stress trough some form of feedback, but \cite{wang_StudentLifeAssessingMental_2014} find that data gathered from smartphones including activity, sleep and conversation data for 48 student participants over 10 weeks forms a lifecycle over the progressing term.
\cite{bogomolov_DailyStressRecognition_2014} collected data from 117 students for one year and tried to predict stress from mobile phone calls, SMS logs, Bluetooth proximity, weather conditions and personality traits, but their stress analysis relies on self-perceived stress level questionnaires answered by the students daily.
\cite{bauer_CanSmartphonesDetect_2012} use an approach similar to ours: They consider a two week exam period as a stressful time and the following two weeks as stress-free and try to distinguish between them based on location traces, Bluetooth device connections, SMS and phone call patterns of 7 participants.
\cite{melillo_NonlinearHeartRate_2011} investigated HRV and a set of HRV features for 42 students during a university exam using short torm ECG. The results show that compared to a unstressed state (after university holidays) almost all HRV features measuring heart rate complexity were significantly decreased during the stressful session.

\subsection{Office Environments}

While office environments are much more diverse in terms of study participants due to age, in literature they are still often studied due to their semi-closeness and ease of gathering data, and are probably the closest form of data acquiring that is not directly university related.
The aim of \cite{vildjiounaite_UnobtrusiveAssessmentStress_2019} was to unobtrusively assess stress of office workers by analyzing their motion trajectories, but they categorize individual participant's days as stressful or not.
\cite{vrijkotte_EffectsWorkStress_2000} monitored 109 male white-collar workers for two work days and one non-work day. They found high stress to be associated with a higher HR during and directly after work, higher blood pressure and lower 24-hour vagal tone, i.e. HRV, irrespective of factors like weight, activity level or caffeine consumption. Additionally, they concluded that values during sleep were more indicative of stress than values during work.
\cite{jarvelin-pasanen_HeartRateVariability_2018} reviews literature dealing with occupational stress and HRV. Of the ten reviewed papers, nine were cross-sectional (i.e., studies with measurements during a single point in time) and one was a longitudinal study with a follow-up after one year. HRV was recorded in up to two between 2 and 36 hour long HR monitoring sessions. Occupational stress was assessed using questionnaires. The review finds a general reduction of HRV and parasympathetic activation with increased occupational stress.
\cite{booth_RobustStressPrediction_2022} explore the efficacy of wearable sensing technologies for stress tracking. By conducting an in situ longitudinal study involving 606 information workers, they leveraged a multimodal approach integrating wearable sensors, relative location tracking, smartphone usage, and environmental sensing in order to predict self-reported stress of participants

As far as we are aware, our combination of dataset, participant number, observed duration and approach of detecting group-wide changes within a rigid curriculum, instead of predicting individual participants' episodes of stress, is unique in the literature.

\section{Data Collection}

\begin{figure}
  \includegraphics[width=1\textwidth]{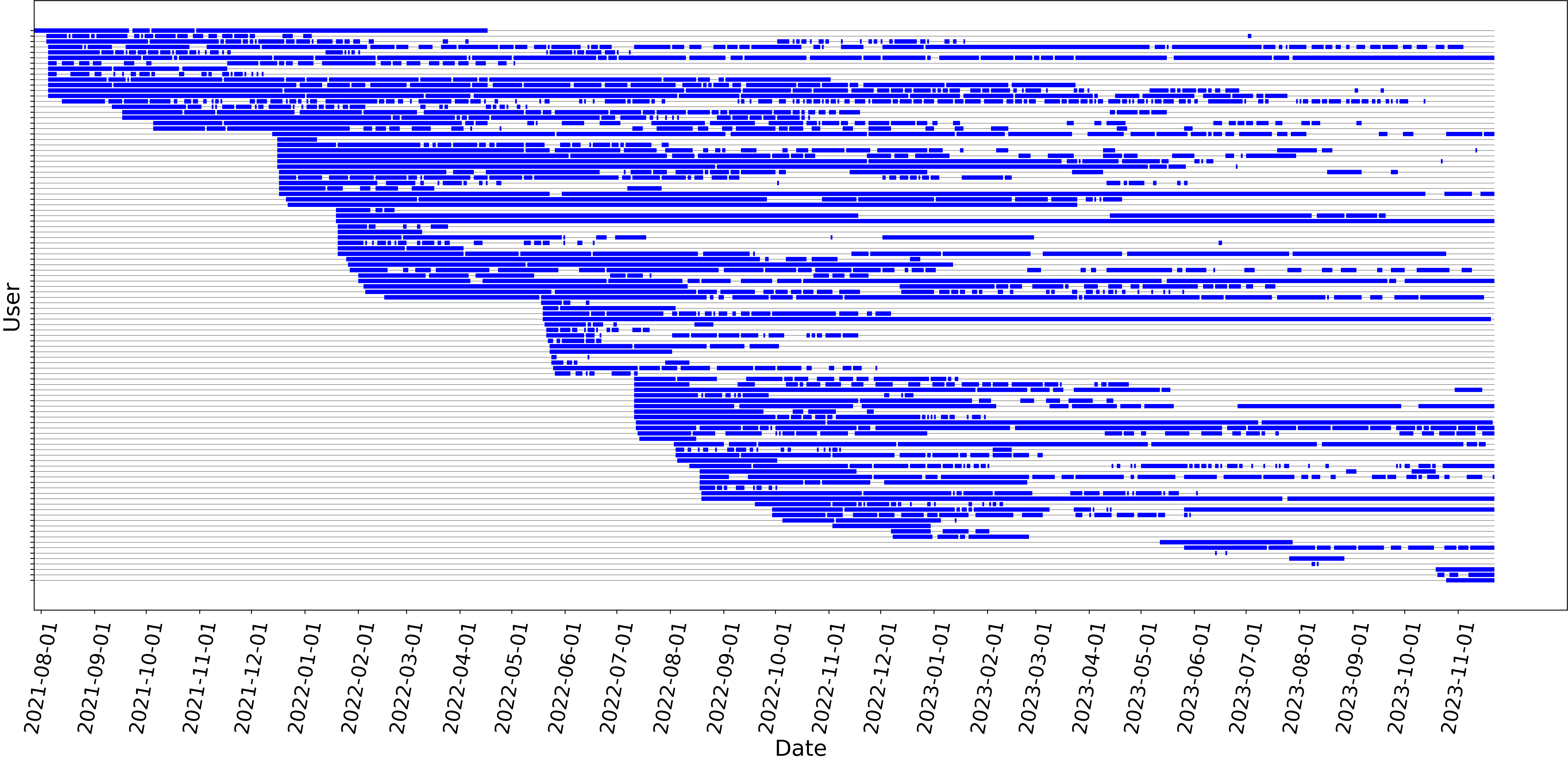}
  \caption{Daily Oura ring data availability (indicative of usage) of participants.}
  \Description{The figure shows a matrix with daily dates ranging from 2021-08-1 to 2023-11-21 on the x-axis and 103 users on the y axis. Entry i-j in the matrix is colored blue if user i has data available for date j. It's visible that users joined the study in different tranches.}
  \label{fig:usage_mat}
\end{figure}

\subsection{Experiment Background}\label{sec:expbackground}
Starting in late 2021, volunteers from a Japanese university have been joining a project on an on-going basis which involves wearing the Oura Ring, a commercial finger-worn sleep and fitness tracker. The participants were graduate and undergraduate university students taking classes affiliated with the engineering school at the host institution. There is no remuneration for wearing the rings, but participants may be remunerated for participating in experiments or answering surveys. The main attraction of the call was the free use of a relatively expensive wearable, as well the promise to have access to their own personal data. There are no special requirements for the participants to wear the ring, and they can leave the project with no penalty. In return for this, the participants agree that their anonymized data can used for research purposes. Because Oura updated their product during the experiment, participants who joined after the beginning of 2022 received updated generation 3 rings instead of generation 2 rings. After receiving their rings, students attended an orientation on the features and capabilities of the Oura Ring and were informed on what data the investigators could collect. In particular, the investigators emphasized that this project is an in-the-wild study and that participants were free to use the rings as they wished. In addition, participants were assured that anonymity of data would be enforced. All data collection and experiments have been approved by the University's Ethics Committee.

\subsection{Choice of Wearable}

There is a wide range of smart health devices available to measure physiological signals related to stress \cite{thapliyal_StressDetectionManagement_2017}. An important factor behind the choice to use the Oura Ring was its form-factor. The Oura Ring is relatively small and unobtrusive, meaning that its presence would not cause a distraction for most participants. At the same time it gives measurements comparable to medical grade devices for sleep HR and HRV~\cite{cao_AccuracyAssessmentOura_2022}. In addition, the battery life requires participants to charge it only once every 4 to 5 days. One full charging cycle takes about 2 hours. Participants have an accompanying smartphone application, where they can see their data. This application needs to be opened one every 4-5 days so that the ring can connect to the application, send the gathered data to the servers and clear its cache. If the application is not opened for longer periods, the ring can overwrite old data which is then lost.

\subsection{How the Oura Ring Collects Data}
The Oura ring automatically collects a range of physiological data during the day using red, green and infrared LEDs as well as an NTC temperature sensor and a 3D accelerometer. The LEDs measure blod vessel dilation \& contraction and blood SpO2 using photoplethysmography, sampling 250 times per second with ``99.9\% reliability compared to a medical-grade electrocardiogram''according to Oura~\cite{_TechnologyOuraRing_2020}. From this measurement HR, HRV and breathing rate are calculated. While the generation 2 rings measure HR and HRV only during sleep, generation 3 rings also measure daytime values. According to Oura (passive) daytime HR measurement is carried out every 5 minutes~\cite{_HeartRateGraph_}, although in practice we find the measuring interval to show fluctuations based on movement~\cite{_TroubleshootingGapsHeart_} and also on the chosen ring size~\cite{neigel_ExploringUsersAbility_2023}. Additionally, generation 3 users can use the smartphone application to get a live measurement of their HR carried out anytime as well as starting workout or meditation sessions, during which the sampling rate for HR is increased. The NTC temperature sensor measures skin temperature every minute, but the temperature data is only published in the form of the deviation of the average temperature during the last night compared to a long term average~\cite{_BodyTemperatureAccessed_}. The 3D accelerometer tracks movements and -- together with HR and HRV values -- is used to classify periods of high/medium/low activity, steps and activity classes. An overview of which data is available for how many participants can be seen in Table \ref{tab:gencomp}.

\subsection{Participants Considered in this Work}
\sethlcolor{cyan}

\begin{figure}
  \includesvg[width=0.6\textwidth]{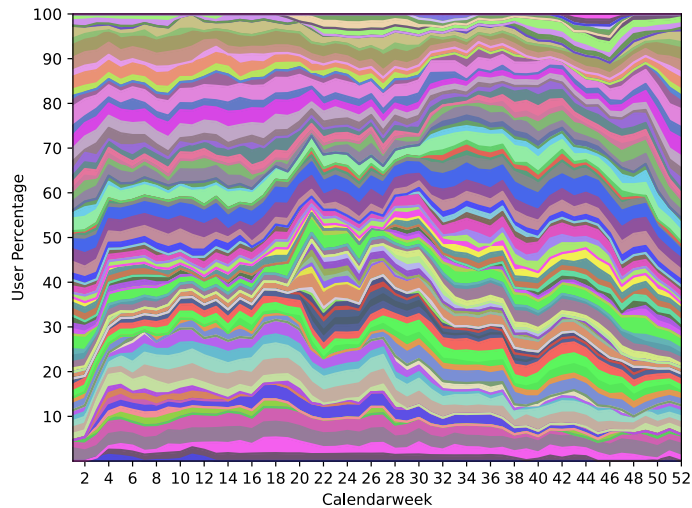}
  \caption{User data distribution per calendar week. Every separated color band corresponds to a single user and the width of the band shows what percentage of the total data available for that calendar week is from that user. No single user or small group of users dominates the data of any given calendar week.}
  \Description{The figure shows how much data for a given calendar week belongs to single users and thereby gives an overview on how balanced the dataset is. It is visible that no user or no small group of users dominate any week.}
  \label{fig:data_dist_calendar}
\end{figure}

Due to the participant onboarding mentioned in section \ref{sec:expbackground}, a total of 151 people have been wearing the Oura ring for up to 28 months. The first and last date considered for analysis are the 30th July 2021 and the 21st November 2023. Since some of the participants are university staff and other non-students and we are interested in student body stress, we considered only students for this study, resulting in $n=103$ participants. Out of these, 91 are male and 12 are female, with mean age $21.8$ (1.9 SD). An overview of the usage of the ring by the participants over that time frame can be seen in Fig. \ref{fig:usage_mat}. Here, the blue marker denotes the availability of either a \textit{daily activity} or \textit{daily sleep} summary for that user and day, indicating usage on that day or the previous night.

To ensure that no single user dominates the data pool during any period, we analyze the share of the total data every user makes up per calendar week. The results can be seen in Fig. \ref{fig:data_dist_calendar}.

\begin{table}[htpb]
\centering
\caption{Oura ring data by generation and percentage of participants.}
\label{tab:gencomp}
\begin{tabular}{lcc}
\toprule
 & Generation 2 & Generation 3 \\
\midrule
Participants & 23 & 80 \\
Sleep HR & \checkmark & \checkmark \\
Sleep HRV & \checkmark & \checkmark \\
Daytime HR & \xmark & \checkmark \\
Daytime HRV & \xmark & \xmark \\
Sleep Breathing Rate & \checkmark & \checkmark \\
Activity & \checkmark & \checkmark \\
Sleep Skin Temperature Deviation & \checkmark & \checkmark \\
\bottomrule
\end{tabular}
\end{table}

\section{Analysis}

\subsection{Academic Year}\label{sec:academicyear}
Universities in Japan usually follow a two-semester system, with the Spring semester starting in April and ending in September and the Fall semester starting in October and ending in March. Final exams are typically given in the first two weeks of August and February respectively. Typically, students enrolled in science and engineering programs will be involved with a laboratory which requires them to accomplish a yearly research project required for graduation.

The time after the semesters' final exam until the start of the next semester is considered as semester breaks. Another period included in our analysis is Golden Week, a week that contains multiple Japanese national holidays and is because of this for students a full week of leisure time that is traditionally used for relaxing and travel~\cite{hara_JapaneseTravelBehavior_2021}.

In addition to studying, students typically begin applying for employment in either their third year or undergraduate studies, or after their first year of graduate studies. Unlike many countries, job-hunting is structured on a strict calendar basis. Job hunting is often stressful for students as it requires them to make trips to companies \cite{kawanishi_ShukatsuUtsuPsychological_2020}.

\subsection{Choice of Physiological Measurements}

As described in section \ref{sec:intro}, stress can manifest itself in a wide array of physiological measurements. In accordance to laboratory studies that analyzed the effects of stress on the body, we look at daytime/waking HR, nighttime/sleep HR and nighttime/sleep HRV as the main indicators of stress. In addition, we look at the daily maximum waking HR of every participant as an indicator of the presence of at least one stressful event during that day. While this daily maximum is more of an indicator of acute stress, sleep HR and HRV are more indicative of lingering or maybe even chronic stress (or drinking the evening before).

Further, we look at sleep related data like total sleep duration, sleep phase percentage, sleep efficiency or restless periods to get further insights.

\subsection{Data Preparation}
\subsubsection{Removal of Seasonal Effects}

\begin{figure}[htpb]
  \includegraphics[width=0.99\textwidth]{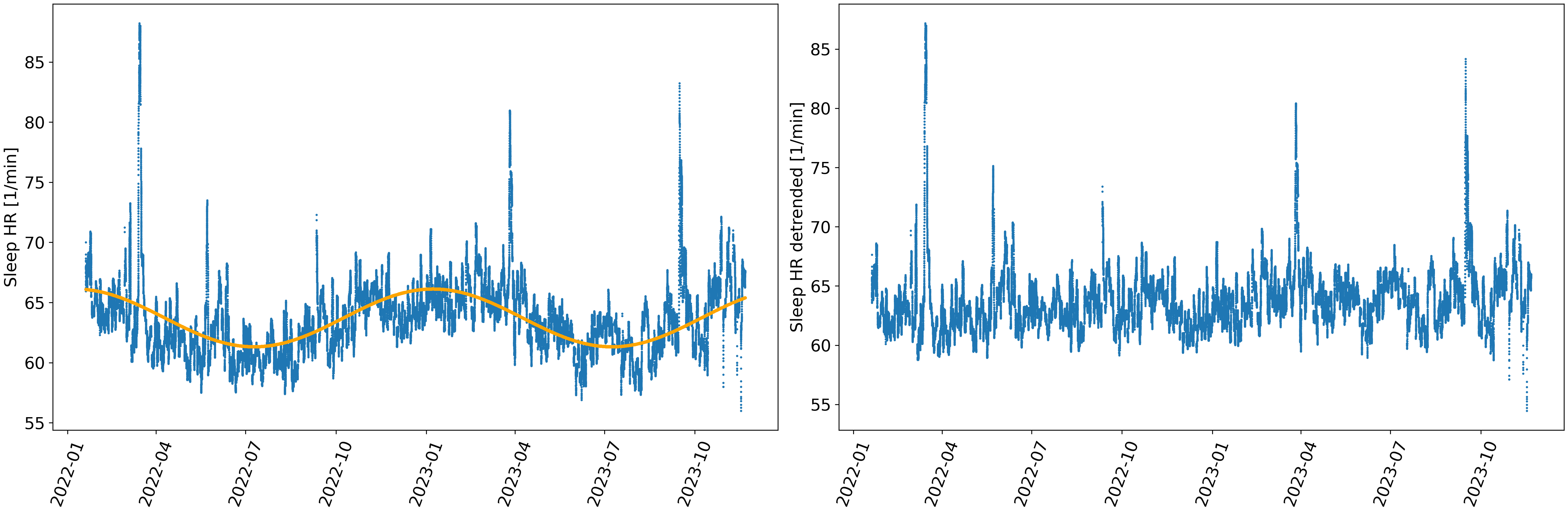}
  \caption{Left: HR recorded during sleep for one participant (blue). The seasonal fluctuation (in unison with daily sunlight hours) is clearly visible. We fit a sinusoidal model (orange) in order to detrend the data. Right: Sleep HR after detrending.}
  \Description{The figure shows two plots, on the left the sleep HR for a user is shown over 22 months, showing a sine-wave-like pattern with peaks  in winter and lows in summer. Right: The same plot but the sine-wave-like pattern is gone due to detrending.}
  \label{fig:detrend-illust}
\end{figure}

As \cite{koskimaki_FollowingHeartWhat_2019} have shown, sleep HR and HRV are subject to a seasonal fluctuations related to daily sunlight hours. We detrend the HR and HRV data captured during sleep by first fitting the data points of every participant to a sinusoidal model:

\begin{equation}
    y = A\cdot\text{sin}(Bt+C)+D \;\;\;\; ,
\end{equation}

where $t$ is the time since the first measurement in seconds, $A$ is the amplitude of the sine wave, i.e. the peak deviation from the central value D, B is the angular frequency determining how many cycles there are in a given period, and C is the phase shift. We fit the parameters to the data by using Maximum Likelihood estimation. From looking at the experimental data and typical seasonal fluctuations, for the optimization we bound parameter A in the range $[0, 5]\frac{1}{\text{min}}$ for HR and $[0,20]\text{ms}$ for HRV. Parameter C was fixed to a value representing a wavelength of 365 days. C was bound to $[-\pi, \pi]$, representing a phase shift between -180 and 180 days, since we consider the first recorded data point as time 0. D was initialized as the average value for that measure over the whole observed period and bounded to the range $[0.5D_{init}, 1.5D_{init}]$ during optimization. We then detrend the data by subtracting the computed model value $y$ and re-adding the vertical shift $D$, effectively removing the seasonal amplitude. An illustration for one participant with the resulting fit can be seen in figure \ref{fig:detrend-illust}.

\subsubsection{Baseline Estimation}

\begin{figure}[htpb]
  \includesvg[width=0.49\textwidth]{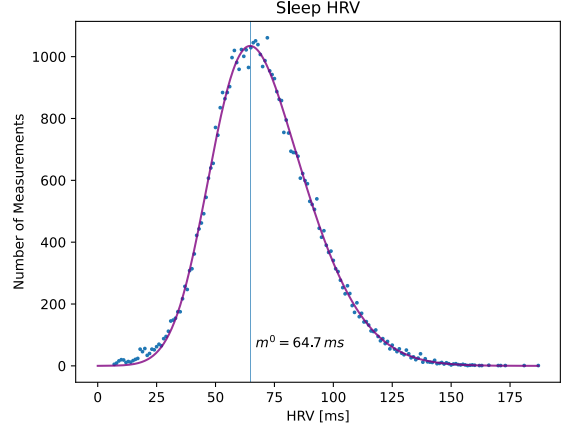}
  \caption{Exemplary illustration of estimation of baseline. A skewed normal distribution is fitted (blue line) to the distribution of a participant's HRV values (scattered dots) from the whole observed period. The argument maximum of the fit, the mode $m^0$, is considered the baseline of that measure (here: HRV) for that participant.}
  \Description{The figure shows a histogram-like scatter plot of HRV measurements. The x-axis shows the HRV value in ms, the y-axis shows how often that value appeared. Also a line is drawn that resembles a skewed-normal distribution that fits the scattered dots very well. The position on the x-axis that shows the highest count is regarded as the baseline.}
  \label{fig:baseline_est}
\end{figure}

Since for measures like HR or HRV baselines differ from individual to individual, in addition to the raw values, we want to normalize all measurements of a user by their respective baseline and variance to ensure comparability in changes between participants. To estimate the baseline for HR and HRV, we model the distribution of these measurements as a skewed-normal distribution, see Eq. \ref{eq:skewednorm} and Fig. \ref{fig:baseline_est}.
\begin{equation}\label{eq:skewednorm}
\begin{split}
    f(x) &= \frac{2}{\omega} \phi\left(\frac{x - \xi}{\omega}\right) \Phi\left(\alpha \left(\frac{x - \xi}{\omega}\right)\right)\\
    \mathrm{where}\;
    \phi(x) &= \frac{1}{\sqrt{2\pi}} e^{-\frac{x^2}{2}}\\
    \mathrm{and}\;
    \Phi(x) &= \int_{-\infty}^{x} \phi(t) dt = \frac{1}{2}\left[1 + \text{erf}\left(\frac{x}{\sqrt{2}}\right)\right]
\end{split}
\end{equation}

Here, $\xi,\omega$ and $\alpha$ are parameters for location, scale and skewness of the distribution, in that order. We estimate these parameters for every measure and user individually by using Maximum Likelihood Estimation.
We then consider the mode of the fitted distribution $m^0$ as the baseline for that measure and that participant. During baseline estimation we ignore HR measurements obtained during live recorded work-outs or guided sessions, since these make up another overlapping distribution with its own parameters, not relevant to the users resting baseline.

As a variance measure, we use the median absolute difference ($MAD$) of measured values to the respective baseline $m^0$:

\begin{equation}
    MAD(M) = \frac{1}{n_{M}}\sum_{x\in M}{(x-m^0_{M})}\;\;\;\; ,
\end{equation}

where $M$ is a type of physiological measure, i.e. either waking HR, sleep HR or sleep HRV, and $n_M$ is the number of measurements a user has for that measure. Finally, we normalize all measures by subtracting the baseline and scaling by the inverse $MAD$:

\begin{equation}
    x' = \frac{x-m^0}{MAD(M)} \;\;\;\; \forall x\in M
\end{equation}

\subsubsection{Daily Aggregation}\label{sec:aggregation}

While some measures supplied by the Oura ring are timestamped by a specific date and describe a summary or aggregation of that measure during that day or the previous night (e.g. total hours slept, total number of steps etc.), other measures like HR or HRV are available as a large number of timestamped measurements (hh:mm:ss) throughout the day or night. For the analysis of the maximum waking HR, we take the daily maximum value per participant, leaving one measurement per participant per (available) day.

\subsection{Mixed Effects Linear Model}

To analyze the physiological data gathered from our participants and its changes during the academic year, we chose to employ mixed effects linear models, because they allow for the modeling of population-wide fixed effects (e.g. periods of the academic year or calendar week) while also taking into account individual variability through random effects.

In mixed effect models, the probability model for a disjoint group \(i\) is:
\begin{equation}
    Y = X\beta + Z\gamma + \epsilon\\ \;\;\;\;\;,
\end{equation}

where $Y$, $X$ and $Z$ correspond to the observed dependent variable, the fixed effects and the random effects respectively, and $\beta$ and the covariance of $\gamma$ are estimated using optimization techniques.

More information about the mixed effects linear model as well as the implementation that we used for our experiments can be found in \cite{seabold2010statsmodels}.

For all our models we used restricted maximum likelihood (REML) optimization to fit the parameters to the model.

\subsection{Pre-Defined Periods}\label{sec:perioddef}

In the first part of the analysis, we cluster data into pre-defined periods based on the academic calendar. The periods, their relevance and exact dates can be found in \ref{tab:predefinedperiods}. The edges of these pre-defined periods are somewhat fuzzy. There could be effects that spill over into one period from another. For instance, the effects of the pre-exam could occur earlier for one student more than another. In addition, not all participants will be experiencing the same influences from the periods as they are happening. Nonetheless, the definitions provide a basis for understanding the status of the cohort as a whole. 

\begin{table}[htpb!]
\centering
\caption{Pre-Defined Periods, their descriptions and dates.}
\label{tab:predefinedperiods}
\renewcommand{\arraystretch}{1.1}
\begin{tabular}{{p{0.15\linewidth}p{0.6\linewidth}p{0.2\linewidth}}}
\hline
    \textbf{Period} & \textbf{Explanation} & \textbf{Dates} \\ \hline
    Spring Exams & Exams end of January, beginning of February. Final exams for the year. & 2021-01-21 until 2021-02-03;
2022-01-20 until 2022-02-02; 2023-01-23 until 2023-02-03\\
   Spring Pre-Exam & The two weeks before the spring exam period. Usually students study for their final exams in this period. & Two weeks prior to the respective spring exam dates \\
    Spring Break & Period after the final exam and before the beginning of the new semester. Many students will be job
hunting in this period, while others may just enjoy leisure activities. & 2021-02-04 until 2021-04-07; 2022-02-03
until 2022-04-07; 2023-02-04 until 2023-04-07\\
    Golden Week & A week that contains multiple Japanese national holidays and is because of this for students a full week of
leisure time & 2021-04-29 until 2021-05-05; 2022-04-29 until
2022-05-05; 2023-04-29 until 2023-05-05 \\
    Summer Exams & Exams end of July, beginning of August, marking the end of the spring semester. & 2021-07-23 until
2021-08-04; 2022-07-22 until 2022-08-04; 2023-07-24 until 2023-08-04  \\
    \makecell[tl]{Summer\\Pre-Exam} & The two weeks before the spring exam period. Usually students study for summer exams in this period. & Two weeks prior to the respective summer exam dates \\
    Summer Break & Period after the summer exams and before the beginning of the autumn semester. & 2021-08-05 until
2021-09-23; 2022-08-05 until 2022-09-23; 2023-08-05 until 2023-09-23\\
 New Year & New Year is an extended holiday, and is typically considered the most important holiday in Japan~\cite{takahashi_IncreasePeopleBehavioural_2022}. & 2021-12-15 until 2022-01-07; 2022-12-15 until 2023-01-07 \\
    Semester & All dates that don’t fall into one of the above periods. This is used as a baseline for most comparisons. & All other dates \\ \hline
\end{tabular}
\end{table}

\section{Results}

\subsection{By Periods}\label{sec:byperiods}

We fit mixed effects linear models with the un-aggregated sleep HR and HRV data (one model per data type), where we set the academic period as a categorical fixed effect and the different users as random effects. We allow for individual intercepts per participant, while keeping a common slope. The results can be seen in table \ref{tab:period-coef-sleep-hr-hrv}.

\begin{table}[htpb]
\centering
\caption{Mixed effects model results for sleep HR and HRV, raw ($\frac{1}{\text{min}}$ \& ms) and normalized values. The leftmost value is the intercept for the reference period (under the semester). All other values indicate the slope coefficients for that period.}
\label{tab:period-coef-sleep-hr-hrv}
\begin{tabular}{l|l|l|l|l|l|l|l|l|l}
\toprule
 & Intercept & \multicolumn{2}{c|}{Semester Break} & & \multicolumn{2}{c|}{Exam}& \multicolumn{2}{c|}{Pre-Exam} & \\
& (Semester) & Spring & Summer & Gol. Week & Spring & Summer & Spring & Summer & New Year's\\
\midrule
Sleep HR \\
\midrule
Raw & 57.07** & 0.57** & 0.59** & -0.23** & -0.47** & 0.28** & 0.47** & 0.27** & 0.34**\\
\multicolumn{1}{r|}{$\pm$} & 0.56 & 0.01 & 0.01 & 0.03 & 0.02 & 0.02 & 0.02 & 0.02 & 0.02 \\
Normalized & 0.77** & 0.19** & 0.21** & -0.06** & -0.13** & 0.11** & 0.17** & 0.09** & 0.11**\\
\multicolumn{1}{r|}{$\pm$} & 0.02 & 0.00 & 0.00 & 0.01 & 0.01 & 0.01 & 0.01 & 0.01 & 0.01 \\
\midrule
Sleep HRV \\
\midrule
Raw & 63.34** & -1.98** & -2.11** & -0.10 & 0.71** & -0.74** & -1.73** & -1.47** & -0.91** \\
\multicolumn{1}{r|}{$\pm$} & 2.23 & 0.04 & 0.05 & 0.12 & 0.08 & 0.09 & 0.09 & 0.08 & 0.06 \\
Normalized & 3.15** & -0.66** & -0.74** & -0.07 & 0.19** & -0.27** & -0.62** & -0.45** & -0.28**\\
\multicolumn{1}{r|}{$\pm$}& 0.87 & 0.01 & 0.02 & 0.04 & 0.03 & 0.03 & 0.03 & 0.03 & 0.02 \\
\bottomrule
\multicolumn{10}{r}{** $p<0.001$, i.e. significant after Bonferroni correction}\\
\multicolumn{10}{r}{Top to bottom: $R^2_{\text{marginal}}=[0.016, 0.041, 0.034, 0.035]$, $R^2_{\text{conditional}}=[0.031, 0.078, 0.068, 0.070]$}\\
\end{tabular}
\end{table}

It can be seen that the general trend and ordering of periods observed is similar for both raw and normalized values. The periods with the highest sleep HR are both semester breaks, the two weeks prior to spring exam and New Year's. The period with the lowest sleep HR is spring exam. For sleep HRV, the period with the highest coefficient is spring Exam and the periods with the lowest values are the semester breaks, and both pre-exam periods. Since higher HR and lower HRV are indicative of stress or recovery, both HR and HRV agree on the three most stressful and two least stressful periods, with some variation in the middle. Only the coefficients for sleep HRV, raw and normalized, during golden week show no significance. At the same time the golden week period shows the lowest coefficients for sleep HR, indicating that the variance between participants is the greatest for this period.

Next, we repeat the procedure for the waking HR. Again, we use the un-aggregated time-stamped heart rates for this model. The results can be seen in table \ref{tab:period-coef-waking-hr}.

\begin{table}[htpb]
\centering
\caption{Mixed effects model results for waking HR, raw ($\frac{1}{\text{min}}$) and normalized values. The leftmost value is the intercept for the reference period (under the semester). All other values indicate the slope coefficients for that period.}
\label{tab:period-coef-waking-hr}
\begin{tabular}{l|l|l|l|l|l|l|l|l|l}
\toprule
 & Intercept & \multicolumn{2}{c|}{Semester Break} & & \multicolumn{2}{c|}{Exam}& \multicolumn{2}{c|}{Pre-Exam} & \\
& (Semester) & Spring & Summer & Gol. Week & Spring & Summer & Spring & Summer & New Year's\\
\midrule
Waking HR \\
\midrule
Raw & 78.12** & 2.31** & 0.88** & 0.11 & 2.04** & 0.60** & 2.30** & -0.02 & 1.54** \\
\multicolumn{1}{r|}{$\pm$} & 1.09 & 0.02 & 0.02 & 0.05 & 0.04 & 0.03 & 0.04 & 0.03 & 0.03 \\
Normalized & 0.73** & 0.29** & 0.10** & 0.01 & 0.26** & 0.07** & 0.28** & -0.01 & 0.19** \\
\multicolumn{1}{r|}{$\pm$} & 0.04 & 0.00 & 0.00 & 0.01 & 0.01 & 0.00 & 0.01 & 0.00 & 0.00 \\

\bottomrule
\multicolumn{10}{r}{** $p<0.001$, i.e. significant after Bonferroni correction}\\
\multicolumn{10}{r}{Top to bottom: $R^2_{\text{marginal}}=[0.091, 0.030]$, $R^2_{\text{conditional}}=[0.394, 0.070]$}\\
\end{tabular}
\end{table}

The periods with the highest coefficients for waking HR are spring break, spring exam and the two weeks prior, followed by new year's. In contrast, summer pre-exam, golden week, the summer exams and summer break show lower values. The order of stressfulness doesn't change when looking at normalized values. Only golden week and summer pre-exam exhibit non-significance, indicating higher variance between participants for these periods.

For the next analysis, we aggregate the waking HR data into daily maximum values, as described in section \ref{sec:aggregation}. This enables us to see the maximum severity of physical stress per day. The results can be found in table \ref{tab:period-coef-waking-hr-max}.

\begin{table}[htpb]
\centering
\caption{Mixed effects model results for maximum waking HR, raw ($\frac{1}{\text{min}}$) and normalized values. The leftmost value is the intercept for the reference period (under the semester). All other values indicate the slope coefficients for that period.}
\label{tab:period-coef-waking-hr-max}
\begin{tabular}{l|l|l|l|l|l|l|l|l|l}
\toprule
 & Intercept & \multicolumn{2}{c|}{Semester Break} & & \multicolumn{2}{c|}{Exam}& \multicolumn{2}{c|}{Pre-Exam} & \\
& (Semester) & Spring & Summer & Gol. Week & Spring & Summer & Spring & Summer & New Year's\\
\midrule
W. HR Max.\\
\midrule
Raw & 119.34** & 0.40 & 1.22** & -1.54 & 1.02 & 3.55** & 0.36 & 2.52** & 0.36 \\
\multicolumn{1}{r|}{$\pm$} & 1.66 & 0.30 & 0.30 & 0.77 & 0.55 & 0.53 & 0.58 & 0.51 & 0.41 \\
Normalized & 5.81** & 0.08 & 0.15** & -0.17 & 0.16 & 0.42** & 0.06 & 0.31** & 0.07 \\
\multicolumn{1}{r|}{$\pm$} & 0.17 & 0.04 & 0.04 & 0.09 & 0.07 & 0.06 & 0.07 & 0.06 & 0.05 \\

\bottomrule
\multicolumn{10}{r}{** $p<0.001$, i.e. significant after Bonferroni correction}\\
\multicolumn{10}{r}{Top to bottom: $R^2_{\text{marginal}}=[0.182, 0.030]$, $R^2_{\text{conditional}}=[0.653, 0.070]$}\\
\end{tabular}
\end{table}

When looking at the maximum waking HR during the day, we find that summer exams and the two weeks prior show the greatest group-wise increase. In contrast, golden week shows the greatest decrease (although with missing significance), followed by New Year's, spring pre-exams and spring break.
Again, raw and normalized values indicate to the same stress-ranking of periods.

\begin{figure}[htpb]
  \includesvg[width=0.95\textwidth]{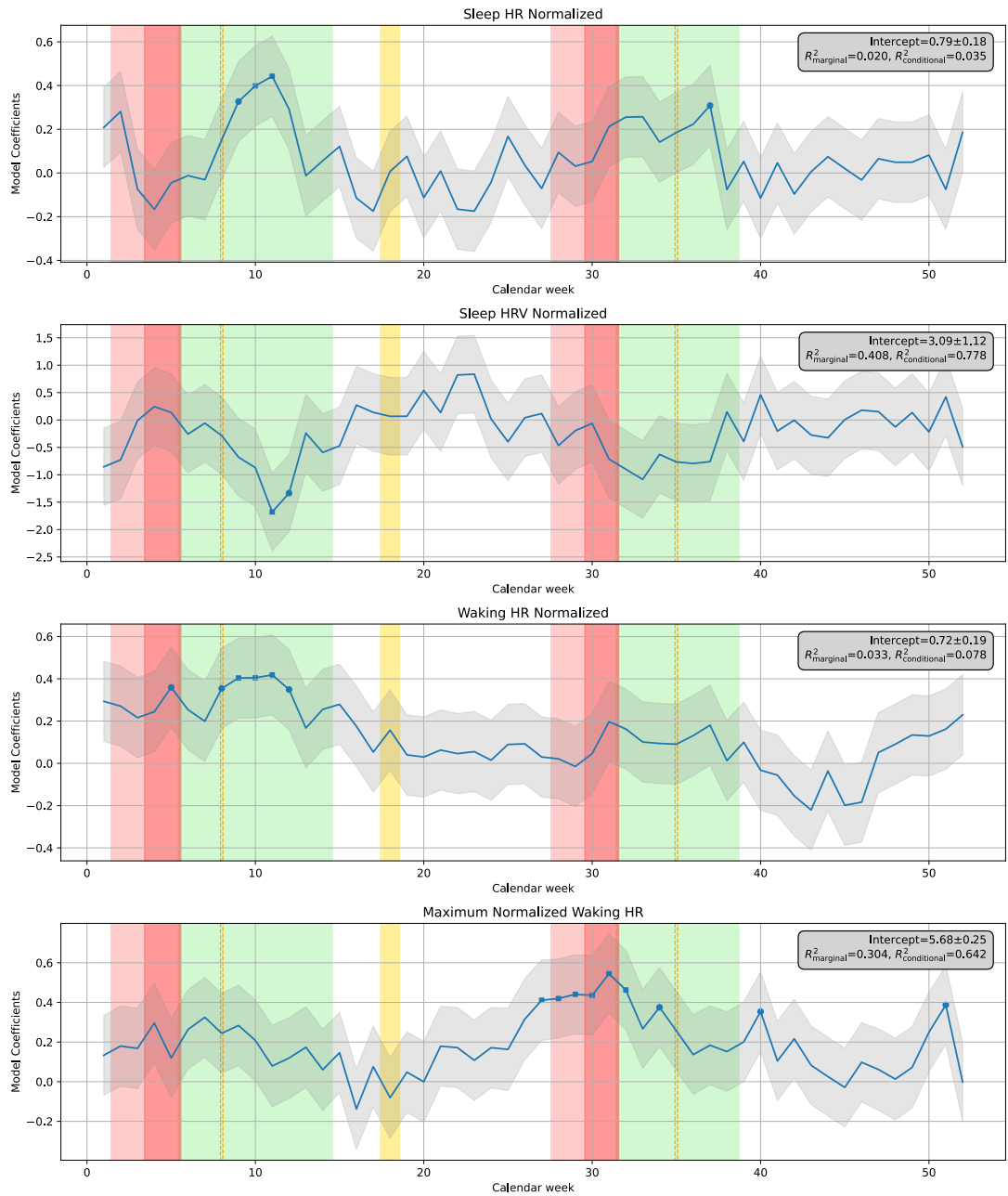}
  \caption{Mixed effects linear model coefficients (unit: MAD) plotted against the respective calendar week for normalized sleep HR, sleep HRV, waking HR and waking HR max. Every graph represents a single fitted model for a single variable denoted at the top of the graph. The background coloring indicates the period according to section \ref{sec:perioddef}: pink - pre-exam, red - exam, green - break, gold - golden week, orange lines - grade result release. The grey box gives the intercept for the reference value -- the median of all values during the semester -- as well as the model fit ($R^2$). Circle dots denote a p-value $<0.1$, square dots a p-value $<0.05$. }
  \Description{The figure shows four plots. Every plot depicts how model coefficients change with calendar weeks, resulting in a curve for the whole year.}
  \label{fig:calweekall}
\end{figure}

\subsection{By Calendar Week}

Since the corner dates of the academic year fall within the same calendar week or within one week difference maximum every year, we chose to "fold" multi-year data by reducing every measurement timestamp to the respective calendar week. Similar to the previous section, we then use the calendar week as a categorical variable that defines the fixed effect of the mixed effects linear model, with the participants as random effects. We allow for individual intercepts per participant but for a common slope. This way, the group-wide movement over the year can be observed visually by plotting the coefficients against the respective calendar week. The reference value for every respective measure to which the denoted intercept value belongs gives the basis from which the plotted coefficients show the change in comparison to that calendar week. For this reference value, we chose to take the median of all measurements that fall into the "semester" category from section \ref{sec:perioddef}.
The resulting coefficients can be seen in figure \ref{fig:calweekall}.

\begin{figure}[htpb]
  \includesvg[width=0.95\textwidth]{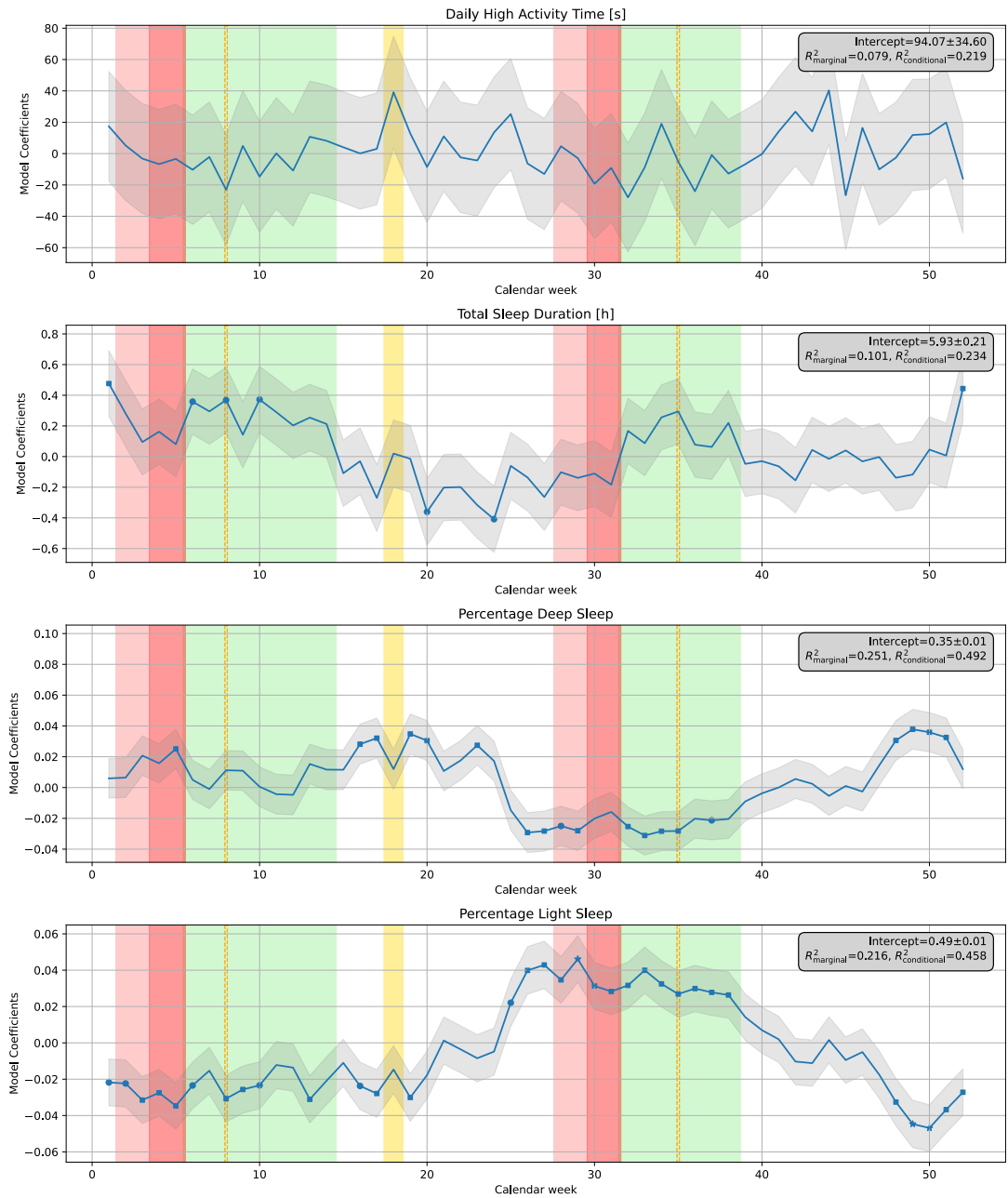}
  \caption{Similar to figure \ref{fig:calweekall}, this plot depicts coefficients for the mixed effects model for the confounding factors high activity time, total sleep duration, percentage of deep sleep and light sleep. Every graph represents a single fitted model for a single variable denoted at the top of the graph. The background coloring indicates the period according to section \ref{sec:perioddef}: pink - pre-exam, red - exam, green - break, gold - golden week, orange lines - grade result release. The grey box gives the intercept for the reference value -- the median of all values during the semester -- as well as the model fit ($R^2$). Circle dots denote a p-value $<0.1$, square dots a p-value $<0.05$. }
  \Description{The figure shows four plots. Every plot depicts how model coefficients change with calendar weeks, resulting in a curve for the whole year.}
  \label{fig:confound}
\end{figure}

Because the considered physiological measurements can vary depending on different influences, but we are mostly interested in stress, we look at further factors related to stress to be able to reason about sources of changes in the measurements. These factors consist of high activity time, measured in seconds per day, total daily sleep duration and percentage of deep and light sleep for the night. We found REM sleep to be relatively constant compared to changes in deep sleep and light sleep and omitted the graph. All these are obtained from the wearable device together with the HR and HRV data. The analysis of these factors is carried out in a similar way to HR and HRV, the results of the mixed effects linear models can be seen in figure \ref{fig:confound}.

Due to less available data per category, many weeks lack the required significance, but the movements of coefficients can give an understanding of patterns nontheless.

\section{Discussion \& Limitations }

\subsection{Discussion of Results}

To interpret these results it is important to distinguish between different types of stress as well as other confounding factors~\cite{schubert_EffectsStressHeart_2009}. While the daily maximum waking HR is indicative of a physiological stressful episode during the day, it's unclear whether that physiological stress is due to a stressful event exciting the participant, e.g. an exam, a presentation, an accident etc., or due to physical activity like sports.
Sleep related values however are indicative of a different type of stress. Sleep HR can increase locally, i.e. for single nights, due to illness~\cite{park_AssociationRestingHeart_2017, davies_RelationshipBodyTemperature_2009, vargo_ConsideringWearableHealth_2023} or alcohol consumption~\cite{pabon_EffectsAlcoholSleep_2022}, but increased HR over several nights or long periods are indicative of chronic stress~\cite{kraaij_RelationshipChronicStress_2020}, fretting and a general need for recovery. Exercise done the day before only has an effect on unaccustomed individuals~\cite{tseng_EffectsExerciseTraining_2020, oconnor_InfluenceTrainingSleeping_1993, koskimaki_FollowingHeartWhat_2019}. Due to the fact that our observed periods consist of several weeks worth of data (every period recorded for a few years), it is reasonable to assume that one-off events like illness can be ruled out, especially considering the significance of most coefficients for the period analysis.

The analysis of waking HR, sleep HR and sleep HRV reveals different periods of elevation (HR) / decrease (HRV) for sleep measures than for waking measures. Sleep HR shows significant group-wide elevation during both spring and summer breaks as well as spring pre-exam and the end of the year, while spring exam and golden week show significant decreases compared to the reference period. Sleep HRV reversely mirrors these movements mostly, showing a rise when sleep HR goes down and vice versa.
Waking HR on the other hand is significantly elevated during the spring months in general: spring exam, pre-exam and break show the highest elevation with new year's following. While the summer exams and summer break also show a significant increase from the reference period, it's not as pronounced as during spring, and it's worth mentioning that the reference period (semester) shows the lowest group-wide HR.
The maximum waking HR has very pronounced and significant increases for both the summer exams and the summer pre-exam period, and to a lesser degree for summer break, indicating stressful events during these periods. Golden week shows a strong decreasing effect on maximum waking HR, but lacking significance.

To further understand the movements leading up to and going out of periods, the model coefficients for single calendar weeks are useful. The strongest indicator for academic related stress is the increase in maximum waking HR just before summer pre-exam period, where many students are beginning to prepare for exams, with a significantly elevated plateau during the pre-exam weeks and significant global peak during summer exams (figure \ref{fig:calweekall}). This is accompanied by a significant reduction in the ratio of deep sleep to the total daily sleep and an increase in light sleep, indicating less restorative sleep and a sign of stress~\cite{kim_EffectPsychosocialStress_2007}, while total sleep duration drops much earlier with the beginning of the semester (figure \ref{fig:confound}). After the summer exams (as well as spring exams), we can see total sleep duration rise sharply, very likely due to the semester breaks starting and students' free time increasing. This increase in maximum HR cannot be explained by heightened activity, as can be taken from the activity coefficients for that period.

While the apparent increase in stress during the summer exams, as indicated by the maximum waking HR, is intuitive, the spring exam period shows a smaller and non-significant increase in the same measure. This is accompanied by a decrease of nighttime HR and increase of HRV as compared to the reference period, showing an apparent contradiction of the stress hypothesis. At the same time, un-aggregated waking HR shows a significant peak during the spring-exam period. This raises the question of how these findings can be mediated. There are some possible explanations that could be speculated in reference to the data, such as the explanation that exams -- and the studying period before -- provide the stressor, as observed by waking HR and maximum waking HR, and the physiological recovery from the stressful period begins after the exams end, during the semester breaks. That sleep HR is increased for both break periods, and especially seems to increase after the grades are released (see figure \ref{fig:calweekall} top) is an indication of this. Another possibility is that the regular schedule that studying for exams provides leads to more restorative sleep. The reality however, is that the analysis in this paper identifies the need for additional data collection, possibly qualitative, to clearly identify the reason for the phenomena.

Another period of relevance is the spring break itself. Waking HR is significantly elevated compared to reference, both for the whole period, but also showing a high plateau in the calendar week analysis. This is accompanied by a rise in HR, a drop in HRV and a drop in deep sleep duration, indicating that the spring semester break is a period of stress for the student body. This seems to harmonize with a peculiarity of the Japanese job market: the job hunting season (section \ref{sec:academicyear}). Further, the spring break is a period of uncertainty for students, besides looking for, applying to and interviewing for jobs, many students often need to find new apartments and can have significant changes to their lifestyles.

Golden week as a period stands out by the fact that for the period analysis only sleep HR yielded significant results with a mild decrease in HR, hinting to a recovery period. That other measures failed to obtain significant coefficients indicates that this week is characterized by a high variation between participants in physiological measurements and possibly behaviours.

The end of the year including new years is again very clear cut: All measures point to a stressful period for the student-body, although the effect is not as pronounced as it is for summer exams. New years in Japan is a major holiday, so students may be traveling or attending parties, which may provide an explanation for the phenomena.

Looking at the maximum waking HR coefficients by calendar week, we see that weeks 40 and 51 show significant increases, indicating that group-wide multi-year stress-events are occurring. The nature of these events are unknown to us, but lead to the broader implications of this study.

\subsection{General Implications}

The results shown in this paper apply to a specific context, students at one university in Japan. This environment is one that contains a lot of structure, benefiting the analyses. The results point to broader implications in regards to the possibilities of group-wide stress detection and implied policy decisions. Our results show that a) it is possible to detect group-wide changes in physiological measurements indicative of broader factors, e.g. stress, without exposing the individual user and their measurements. A person's health data is considered as private and security sensitive, giving greater importance to this finding. Moreover, even though our cohort is homogeneous compared to the general population, this cohort could be split into more fine-grained groups to improve stress detection in periods. The distinction by current academic year for example would give an insight into which students are writing theses or doing laboratory research during breaks, or which groups have important exams in a specific semester, indicating higher stress. On an individual level, one could differentiate between students showing no or less signs of stress during exams due to them being very confident in their abilities and stressed students. That it is possible to detect these group-wide patterns without these fine-grained subdivisions strengthens the general applicability and maintains user privacy. In addition, our findings show that b) the usage of the wearable device doesn't need to be strictly enforced. In our study participants have been wearing the ring as they saw fit, skipping many days and, in some cases, stopping usage of the ring completely. Even with this kind of natural in-the-wild data, which is messy and has gaps, it is possible to detect group wide patterns.

The implications of this study are therefore likely applicable to other types of groups in a wide-arrange of contexts. This is important, because it opens the opportunities for the usage of wearable sensors, like the Oura Ring, for the purpose of policy making and management at large institutions which do not seek or have control over their potential wearers. This is important, since we know that devices like the Oura Ring have great potential in contexts where control is maintained, such as the military~\cite{conroy_RealtimeInfectionPrediction_2022}. In the university context, having relatively few volunteers wear a device like the Oura Ring could help administrators recognize unknown periods of stress, where they could then react by changing curriculum structures or preparing other ways to mitigate adverse campus trends. This is a opposite direction from where universal correct usage of the device is required in order for the device being worth deploying. It is also more feasible for an institution to achieve, from both a device deployment aspect, as well as from an infrastructure development and maintenance aspect. 

\subsection{Limitations}

One of the limitations of our study is the relative lack of female participants in our cohort. While the uniformity of our participants in terms of age, student status, and adherence to the academic calendar of a Japanese university are all points we consider strengths due to the fact that it enables the relevant policymakers (e.g., university deans) to consider actions specifically for this group, the lack of female participants is a limitation on the generalizability of the results, especially considering the effect of gender on the relationship between stress and HR~\cite{kraaij_RelationshipChronicStress_2020}.

We also only look at just physiological data, without taking into account contextual data like movement patterns, smartphone use, sociability etc., which were shown to improve stress detection over just physiological data~\cite{stojchevska_AssessingAddedValue_2022}. While this is a limitation in one sense, these added data points would increase complexity and the amount of data that is required from each user, thus creating increased privacy risks.

On an analysis level, we only use a single fixed effect for our mixed effect models. It might be possible to include a more detailed analysis, the interaction between periods, confounding and explaining factors like activity or sleep and biomarkers of stress could shed light on stress and their causes within the cohort.

Finally, we do not expand this study into looking for the reasons for stress in the cohort. While the academic calendar allows for us to make reasonable inferences about the causes, there may be underlying reasons that are unseen from our approach alone. Likely, qualitative investigations are needed to find any underlying causes. However, our work does uncover the need for such investigations. 

\section{Conclusion \& Future Work}

In this paper, we sought to detect stress in-the-wild by analyzing the HR, HRV, sleep, and activity data of 103 participants gathered from a wearable health ring for up to 28 months, for the periods of summer exams, spring break, and New Year's' clear signs of elevated physiological stress were identified. While the first two are very intuitive, a possible explanation for the latter can be the Japanese job market and rigid academic schedule, leading to many lifestyle changes for students. We conclude, therefore, that the distribution of wearable health devices, especially rings, in relatively small numbers allows for unobtrusive, privacy-preserving analysis of group-wide effects, without the need to enforce usage and despite the fact that the data is noisy. This can also enable the identification of previously unknown forms of stress or other group-wide effects, having broad implications for universities' policies. To improve this research direction, future work could add ways to prompt users about stressful events, ideally daily. This way, physiological measurements could be compared to participants' self-reflections, possibly strengthening the periodic analysis. Taking into account contextual information like movement patterns, smartphone use, sociability, etc. could further improve stress-detection capabilities.

\begin{acks}
This work was supported in part by grants from the
JST Trilateral AI Research (JPMJCR20G3), JSPS Grant-in-Aid for Scientific Research (20KK0235, 23KK0188) and
the Grand challenge of the Initiative for Life Design Innovation (iLDi)
\end{acks}

\bibliographystyle{splncs04}
\bibliography{bib}
\end{document}